\documentclass[preprintnumbers,twocolumn,superscriptaddress,
floatfix,showpacs,prc,aps,amsmath,amssymb,nofootinbib]{revtex4-1}
\usepackage{graphicx}

\usepackage{hyperref}
\usepackage{multirow}
\usepackage{adjustbox}
\usepackage{caption}

\begin{document}

\title{Freeze-out configuration properties\\
 in the $^{197}Au + ^{197}Au$ reaction at 23 AMeV}

\author{R.Najman}
\affiliation{M.Smoluchowski Institute of Physics, Jagiellonian University, Krak\'ow, Poland}

\author{R.P\l{}aneta}
\email[E-Mail:]{roman.planeta@uj.edu.pl}
\affiliation{M.Smoluchowski Institute of Physics, Jagiellonian University, Krak\'ow, Poland}

\author{A. Sochocka}
\affiliation{Department of Physics, Astronomy and Applied Informatics, Jagiellonian University, Krak\'ow, Poland}

\author{F.Amorini}
\affiliation{INFN, Laboratori Nazionali del Sud, Catania, Italy}
\affiliation{Dipartimento di Fisica e Astronomia Universit\`a di Catania, Catania,Italy}

\author{L.Auditore}
\affiliation{Dipartimento di Fisica e Scienze della Tera Universit\`a di Messina and INFN Gruppo Collegato di Messina, Italy}

\author{T.Cap}
\affiliation{Faculty of Physics, University of Warsaw, Warsaw, Poland}

\author{G.Cardella}
\affiliation{INFN, Sezione di Catania, Italy}

\author{E. De Filippo}
\affiliation{INFN, Sezione di Catania, Italy}

\author{E.Geraci}
\affiliation{Dipartimento di Fisica e Astronomia Universit\`a di Catania, Catania,Italy}
\affiliation{INFN, Sezione di Catania, Italy}

\author{A.Grzeszczuk}
\affiliation{Institute of Physics, University of Silesia, Katowice, Poland}

\author{S.Kowalski}
\affiliation{Institute of Physics, University of Silesia, Katowice, Poland}

\author{T.Kozik}
\affiliation{M.Smoluchowski Institute of Physics, Jagiellonian University, Krak\'ow, Poland}

\author{G.Lanzalone}
\affiliation{INFN, Laboratori Nazionali del Sud, Catania, Italy}
\affiliation{Universit\`a degli Studi di Enna ``Kore'',Enna, Italy}

\author{I.Lombardo}
\affiliation{Dip. di Fisica, Universit\`a di Napoli Federico II, Naples, Italy}
\affiliation{INFN, Sezione di Napoli, Italy}

\author{Z. Majka}
\affiliation{M.Smoluchowski Institute of Physics, Jagiellonian University, Krak\'ow, Poland}

\author{N.G.Nicolis}
\affiliation{Department of Physics, The University of Ioannina, Ioannina, Greece}

\author{A.Pagano}
\affiliation{INFN, Sezione di Catania, Italy}

\author{E.Piasecki}
\affiliation{Heavy Ion Laboratory, University of Warsaw, Warsaw, Poland}
\affiliation{National Centre for Nuclear Research, Otwock-\'Swierk, Poland}

\author{S.Pirrone}
\affiliation{INFN, Sezione di Catania, Italy}

\author{G.Politi}
\affiliation{Dipartimento di Fisica e Astronomia Universit\`a di Catania, Catania,Italy}
\affiliation{INFN, Sezione di Catania, Italy}

\author{F.Rizzo}
\affiliation{INFN, Laboratori Nazionali del Sud, Catania, Italy}
\affiliation{Dipartimento di Fisica e Astronomia Universit\`a di Catania, Catania,Italy}

\author{P.Russotto}
\affiliation{INFN, Sezione di Catania, Italy}

\author{K.Siwek-Wilczy\'nska}
\affiliation{Faculty of Physics, University of Warsaw, Warsaw, Poland}

\author{I.Skwira-Chalot}
\affiliation{Faculty of Physics, University of Warsaw, Warsaw, Poland}

\author{A.Trifiro}
\affiliation{Dipartimento di Fisica Universit\`a di Messina and INFN Gruppo Collegato di Messina, Italy}

\author{M.Trimarchi}
\affiliation{Dipartimento di Fisica Universit\`a di Messina and INFN Gruppo Collegato di Messina, Italy}

\author{J.Wilczy\'nski}
\affiliation{National Centre for Nuclear Research, Otwock-\'Swierk, Poland}

\author{W.Zipper}
\affiliation{Institute of Physics, University of Silesia, Katowice, Poland}

\date{\today}

\begin{abstract}
Data from the experiment on the $^{197}Au + {}^{197}Au$ reaction at 23 AMeV are analyzed with an
 aim to find signatures of exotic nuclear configurations such as toroid-shaped objects.
 The experimental data are  compared with predictions of  the ETNA code dedicated to look for
 such configurations and with the QMD model. A novel criterion of selecting events possibly resulting
 from the formation of exotic freeze-out configurations, "the efficiency factor", is tested. Comparison
 between experimental data and model predictions may indicate for the formation of flat/toroidal nuclear systems.
\end{abstract}

\maketitle

\section{Introduction}

The search for exotic nuclear configurations was inspired by
J.A.Wheeler \cite{Wheeler50}. His idea was investigated by many authors who
studied the stability of exotic nuclear shapes (see e.g. \cite{Siemens67, Wong85, Moretto97}).
Theoretical investigations have shown that very exotic extra superheavy nuclei can be reached
only if non-compact shapes are taken into account. Calculations for bubble structures
showed that such nuclei can be stable for $Z>240$ and $N>500$ (see e.g.
\cite{Dietrich98, Berger01, Decharge03}). Recently it was found that for nuclei with $Z>140$ the
global energy minimum corresponds to toroidal shapes \cite{Warda07, Staszczak09}. In contrast
to bubble nuclei, the synthesis of toroidal nuclei is experimentally available
in collisions between stable isotopes.

To address this issue simulations were performed for Au + Au collisions in a wide range of
incident energies using the BUU code \cite{Sochocka08, Sochocka108}. These
calculations indicate that the threshold energy for the formation of toroidal
nuclear shapes is located around 23 AMeV.

Also Improved Quantum Molecular Dynamics Model calculations performed for U +
 U collisions have shown a possible formation of toroidal freeze-out configurations
above a specific collision energy for this heavy system
\cite{Tian08}. Such toroidal-shape complex can be also created in macroscale in
binary droplet collisions above some threshold velocity \cite{Kuo09}.

A number of observables were suggested as the signatures of noncompact
freeze-out configurations. These were:

 \begin{itemize}
 \item Larger number of intermediate mass fragments should be observed than
 would be expected for the decay of a compact object;
 \item Enhanced similarity in the charge and size of the fragments should
 also be observed;
 \item Suppressed sphericity in the emission of fragments should be
 visible.
 \end{itemize}

The simulations of decay process of different break up configurations using the
ETNA code were performed to study the ability of the CHIMERA detector
\cite{Pagano04, Defilippo14} for recognition of non-compact configurations. Analysis of different
observables have shown that a quantity named ``the efficiency factor'' of
events with 5 heavy fragments can be used as a criterion of selecting events possibly resulting
from formation of toroidal configurations \cite{Sochocka09, Sochocka09phd}.

 The CHIMERA collaboration has carried out an experiment on the $^{197}Au + {}^{197}Au$ reaction at 23 AMeV beam energy,
 focused on two independent goals, first on the extension of the earlier study at lower energy of 15 AMeV,
  in which a new reaction mechanism of violent breakup of non-fusing $^{197}Au + {}^{197}Au$ system into 3 and/or 4 massive
 fragments was observed \cite{Skwira08}, \cite{Wilczynski10}, \cite{Wilczynski10a}, and second, on the search of exotic nuclear configurations such as toroidal shapes.
 Some preliminary results of the former project have been published in \cite{Cap14}.

 In this work we report results of our analysis focused on the question of exotic configurations involving the
 breakup of the $^{197}Au + {}^{197}Au$ system into 5 or more fragments. The experimental data are compared with model predictions.
 Conclusions regarding the shape of the freeze-out configuration are drawn.

This paper is organized as follows. In Sec. 2 we present the experiment and data calibration procedures.
General characteristics of experimental data are shown in Sec. 3. The dedicated observables are discussed in Sec. 4.
The conclusions are presented in Sec. 5.

\section{Experiment and data calibration procedure}

The experiment for the $^{197}Au + {}^{197}Au$ reaction at 23 AMeV was performed at INFN-LNS Superconducting
Cyclotron of Catania. During the experiment two gold targets
 were used: 164 and 396 $\mu
g/cm^{2}$. The thinner target was used in calibration measurements and the
thicker one in the production runs. Reaction products were detected 
with the CHIMERA multidetector \cite{Pagano04, Defilippo14} that is constituted by 1192 telescopes arranged 
in 35 rings in full $2\pi$ azimuthal symmetry around the beam axis, covering the polar angle
between $1^{o}$ and $176^{o}$. A single detection cell is constituted of a planar n-type silicon 
detector  ($\cong 300$ $\mu$ thickness) followed by a Cs(Tl) scintilator of thickness
varing from 12 cm at forward angles to 3 cm at backward angles.

The collected data were calibrated using a set of dedicated programs developed at
INFN-LNS. Energy calibration of Si detectors was performed using ion beams, delivered both
by the tandem and the cyclotron. Data for the following systems were used: (i)
the elastic scattering data for $^{16}O + Au$ at 60 and 80 MeV, $^{58}Ni + Au$
at 142 MeV, Au + Au at 170 MeV and 23 AMeV; (ii) recoil peak for $Au + ^{12}C$
at 170 MeV; and (ii) fission fragments from $Au + ^{12}C$ reaction at 23 AMeV.
Unlike in the analysis of ternary breakup reactions \cite{Cap14} in which the pulse-height defect was calculated
 with the formula of Ref. \cite{Tabacaru}, 
 in the present analysis the pulse-height defect in silicon detectors was calculated using the same procedure
 as described in Ref. \cite{Pasquali98}.

\begin{figure}[t]
 \includegraphics[width=0.44\textwidth]{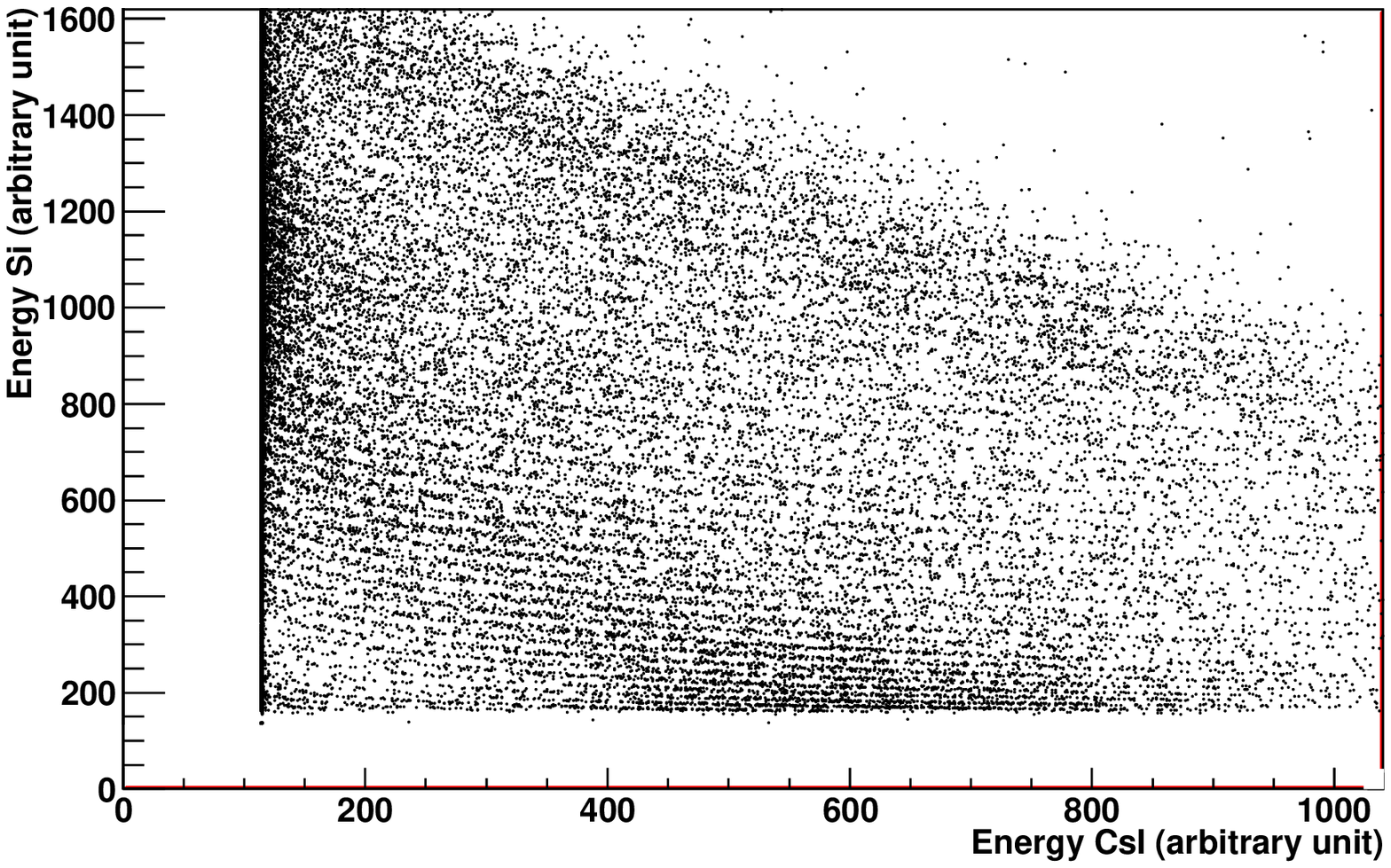} 
 \hspace{1.5cm}
\includegraphics[width=0.39\textwidth]{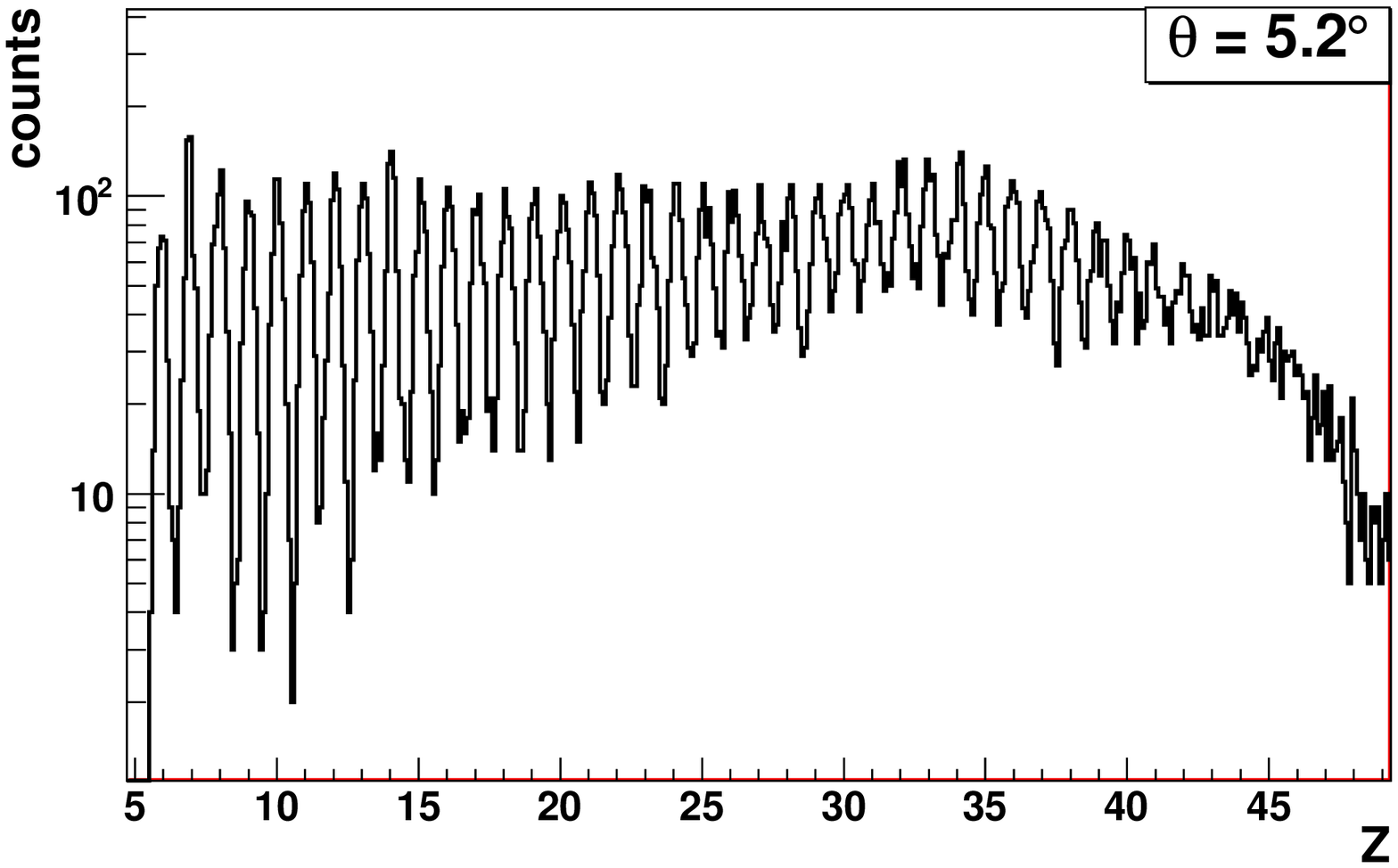}
\caption{$\Delta E-E$ spectrum(upper panel) and the corresponding Z spectrum (bottom panel) for fragments detected
in telescope placed at $\theta = 5.2^{o}$ for Au + Au reaction at 23 AMeV.}
\label{fig01}
\end{figure}

In order to identify fragments two methods were applied: (i) the $\Delta E-E$ technique for fragments punching
through the silicon detectors; (ii) the time of flight (TOF) method for the class of fragments stopped in Si detectors.
In Fig.~\ref{fig01} (upper part) an example of $\Delta E-E$ plot is shown
for a detector belonging to the 3-th internal ring, at a polar angle
$\theta=5.2^{o}$. The Z distribution of fragments identified by the $\Delta E-E$ method for the same detector
is presented also in Fig.~\ref{fig01} (bottom panel). We can see that in the Z spectrum a good charge
identification can be observed up to $Z=42$. At this angular range one observes a broad maximum of the charge distribution
located at charges $Z~=~30-40$ corresponding to Au fission fragments. One observes also
a substantial contribution of lighter fragments.
In order to estimate a missing information on mass of the fragments identified in charge by $\Delta E-E$
method, the EPAX formula \cite{epax1, epax2,epax3} was used.

\begin{figure}
\begin{minipage}{0.95\linewidth}
\begin{center}
\includegraphics[width=0.9\textwidth]{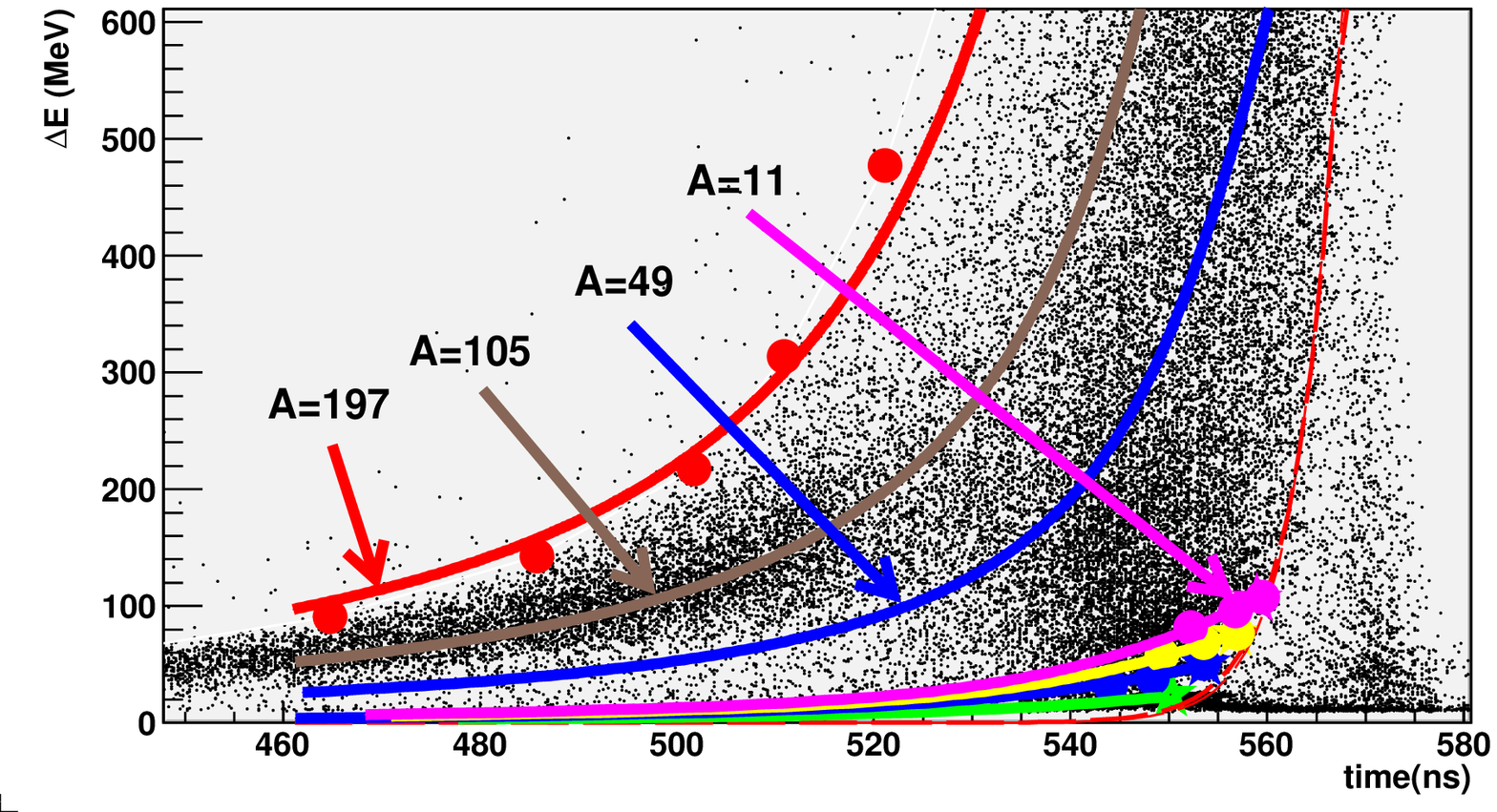}
\includegraphics[width=0.9\textwidth]{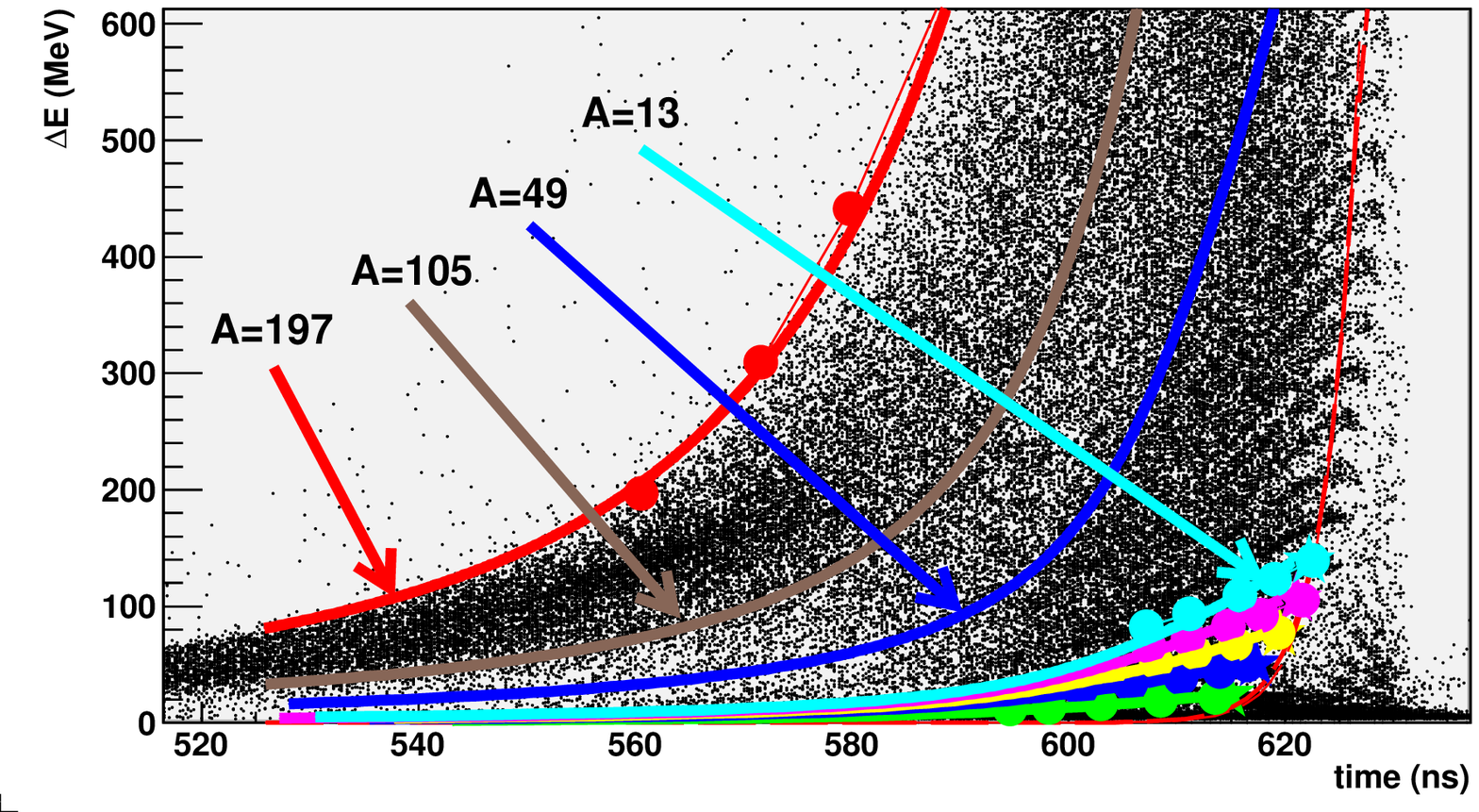}
\end{center}
\end{minipage}
\caption{
(color online) The $\Delta E-TOF$ distributions for detectors 404 (upper panel) and 653 (bottom panel) located at $\theta = 17^{o}$ and $28.5^{o}$, respectively.
Lines presented of both figures correspond to positions of masses as indicated.
}
\label{fig02}
\end{figure}

 The mass of fragments stopped in Si detector is determined by TOF method. The start signal was given
 by 30\% Constant Fraction Discriminator acting on time signal generated
 by the silicon detector, while the stop signal was given by delayed Reference Signal delivered by cyclotron.
 Examples of $\Delta E-TOF$ spectra are presented in Fig.~\ref{fig02}.
 In this case mass values are calculated using the formula:

\begin{equation}
m=2E\cdot(t_{0}-t)^{2}/R^{2},
\label{P1}
\end{equation}%

where R is the distance between the target and a given detector and the $t_{0}$ is a time offset of the measured time t. 

\begin{figure}
\begin{minipage}{0.99\linewidth}
\begin{center}
\includegraphics[width=0.99\textwidth]{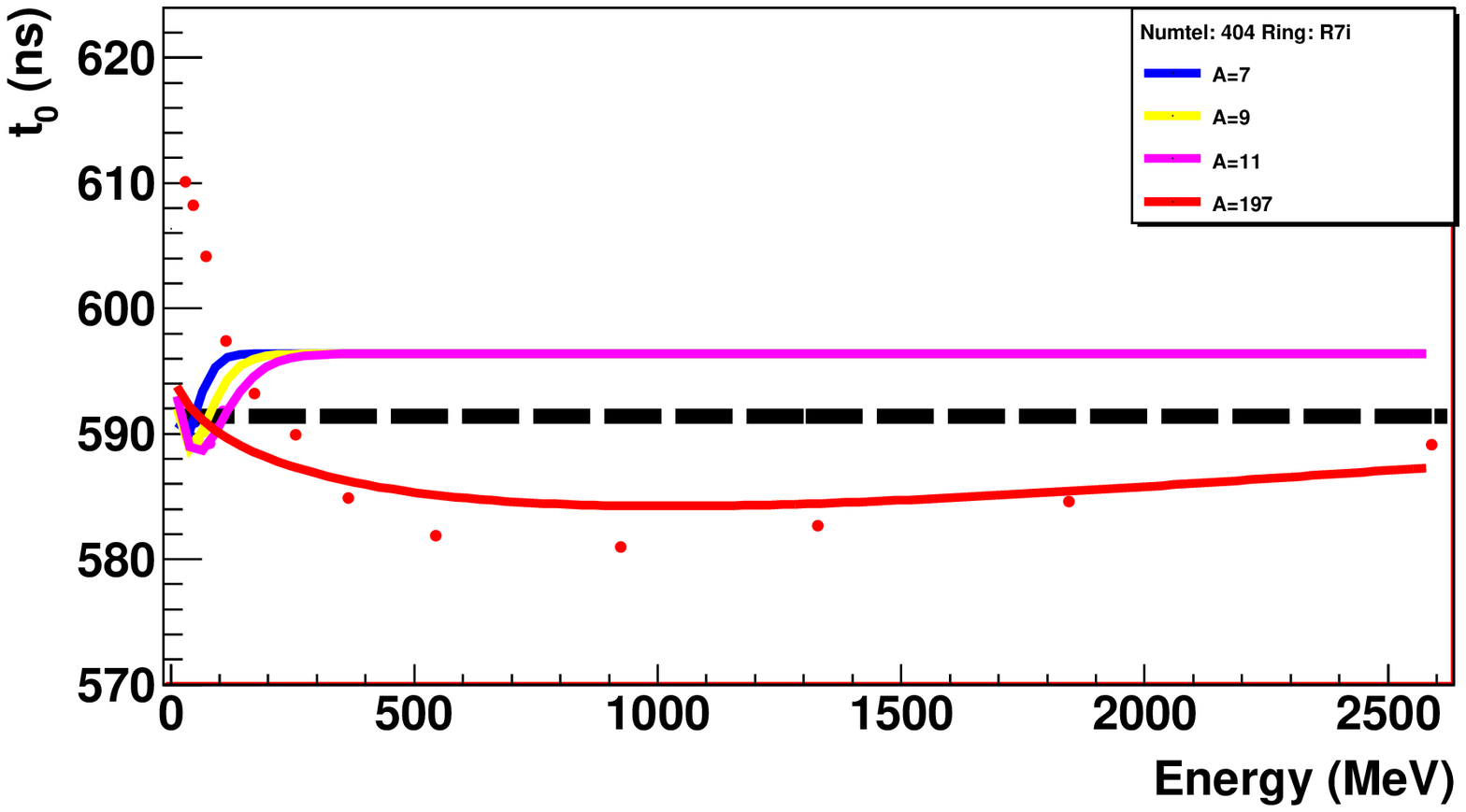}
\includegraphics[width=0.99\textwidth]{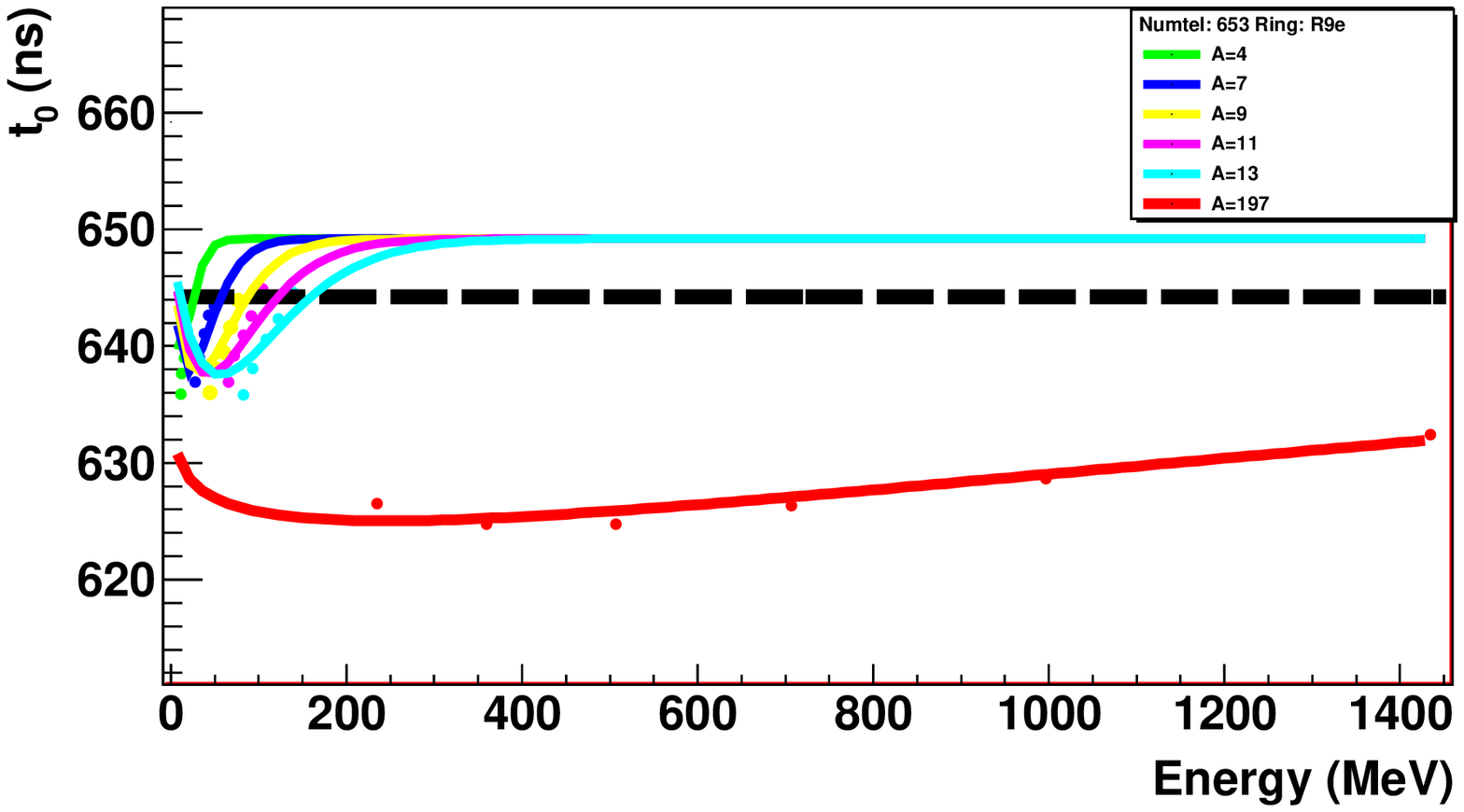}
\end{center}
\end{minipage}
\caption{(color online) The $t_{0}$ dependence on incident energy and particle mass for telescopes 404 and 653 located at
$\theta = 17^{o}$ and $28.5^{o}$, respectively. Color symbols represent the $t_{0}$ values
for identified light fragments and Au-like fragments.
Solid lines represent the fitting results using the formula. The dashed line indicate the $t_{0,sat}$ value. 
}
\label{fig03} 
\end{figure}

A crucial problem in the calibration of TOF measurements for the CHIMERA multidetector is evaluation of
  $t_{0}$  offset that must be determined for each detector individually. Moreover, $t_{0}$  depends on mass,
 charge and kinetic energy of the detected fragment. The $t_{0}$ values for well identified light fragments and Au-like nuclei
 fragments located at the left-hand-side edge of the $\Delta E-TOF$
 distribution (see Fig.~\ref{fig02}) are presented by color symbols in Fig.~\ref{fig03}.
 For relatively light fragments a well
 tested parametrization of $t_{0}$ for CHIMERA detectors was proposed \cite{Lanzalone}.
 To calibrate $t_{0}$ for medium and heavy fragments in $^{197}Au + {}^{197}Au$ experiments a new 
calibration method based on a functional dependence of $t_{0}$ on mass, charge and pulse-height-defect
 dependent kinetic energy was developed (see e.g. \cite{cap13})  and
 applied in analysis of  the ternary breakup experiment \cite{Cap14}.

In the present analysis we use another method of the parametrization of $t_{0}$ offset:

\begin{equation}
t_{0}=	 \begin{cases}
 t_{0,sat} & t_{0,sat} < \Delta t \\
 t_{0,sat}-\Delta t & t_{0,sat} > \Delta t
 \end{cases}
\end{equation}

\begin{eqnarray}
\label{tt0}
 \Delta t= B-A(1-exp(\gamma \cdot m)) \cdot (\frac{E}{E_{PT}})^{(\alpha-\delta \cdot m)}  \nonumber \\
 \cdot exp[-(\frac{E+(\beta+ \mu \cdot m) E_{PT}}{E_{PT}})^{\epsilon}],
\end{eqnarray}

where $t_{0,sat}$ is determined for particles punching through the silicon detector. The $E_{PT}$ is the highest energy deposited by
particles with mass m. 

The values of the parameters A, B, $\alpha, \beta, \gamma, \mu$ and $\epsilon$ were determined by fitting Eq. (2)
 to selected points  for well identified  light fragments and Au-like nuclei
 fragments (see Fig. ~\ref{fig03}). In this procedure fragment energies were corrected for pulse-height defect and their Z values were estimated
 using the EPAX formula [22-24].
 Fig. ~\ref{fig03} presents the $t_{0}$ calibration lines calculated  for selected mass values.
  The calculated positions in the $\Delta E - TOF$ distributions for some selected
 mass numbers are shown for two detectors in Fig. ~\ref{fig02} as solid color lines.

\begin{figure}
\begin{center}
\includegraphics[width=\columnwidth]{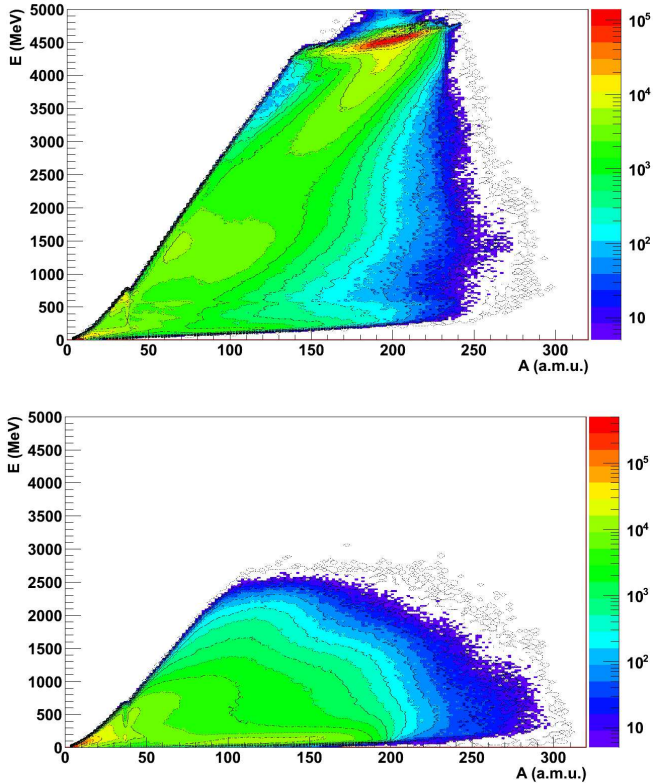}
\caption{(color online) The correlation between mass and energy for fragments observed in telescopes located at
 $3^{o} < \theta < 10^{o}$ (upper panel), and $20^{o} < \theta < 29^{o}$ (bottom panel).}
\label{fig04} 
\end{center}
\end{figure}

In Fig.~\ref{fig04} two dimensional mass versus kinetic energy distributions are shown for telescopes located at two angular regions.
 For $3^{o} < \theta < 10^{o}$ region (upper panel) the distribution extends from small masses seen at low energies up to the Au elastic peak.
For $20^{o} < \theta < 29^{o}$ region (bottom panel) the particles with masses up to 200 a.m.u. are observed at relatively low kinetic energies.

\section{The general characteristics of experimental data}
\begin{figure}
\begin{center}
\includegraphics[width=\columnwidth]{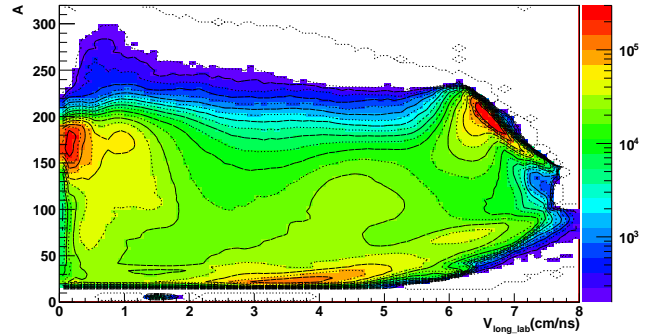}
\caption{(color online) The correlation between mass of identified
fragments versus parallel velocity of those fragments.}
\label{fig05} 
\end{center}
\end{figure}

In Fig.~\ref{fig05} two dimensional distribution mass versus parallel velocity of identified fragments is shown. Location of quasielastic Au peak is visible
at mass around 200 and velocities close to the beam velocity ($v_{p} = 6.67 cm/ns$). Peak corresponding to
Au recoil fragments can be found at velocities close to zero. Here one can observe an
underestimation of mass value for these fragments due to the imperfection of  used $t_{0}$ parametrizations (see Eq. 2).
At velocities between these
two limits fragments originating from fission of the Au-like nuclei are located. One can also identify a separated region located at low masses
 and velocity close to center of mass velocity. This region correspond to the intermediate velocity source.

\begin{figure}
\begin{center}
\includegraphics[width=\columnwidth]{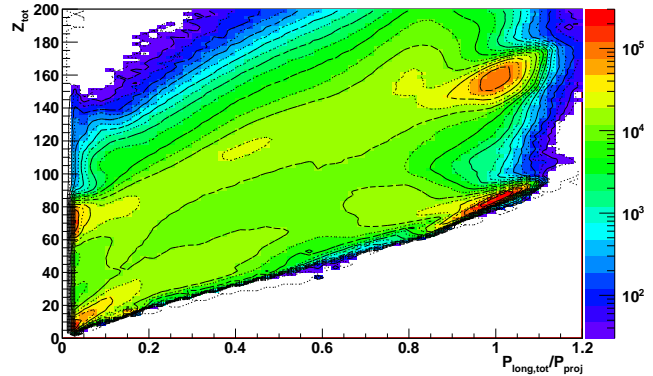}
\caption{(color online) The correlation between total charge of identified
fragments versus total parallel momentum of those fragments normalized to the beam momentum.}
\label{fig06} 
\end{center}
\end{figure}

For the registered events we have constructed the plot presenting the
dependence between the total charge of identified fragments, $Z_{tot}$, versus total
parallel momentum of those fragments normalized to the beam momentum, $p_{\|,tot}/p_{proj}$ (see Fig.~\ref{fig06}).

One can distinguish different regions on this plot. In the region of low values of total collected charge
and low paralled momentum one observes the ridge corresponding to the badly detected events. In the region
of total parallel momentum close to 1 and total collected charge close to the charge of projectile one observes
the maximum corresponding to deep inelastic collisions where the target like fragment remains undetected.
Region where the total detected charge is close to total charge of the system and the total parallel linear momentun is
close to linear momentum of the projectile can be called as region of well reconstructed events.
 In our present analysis this region is selected imposing the conditions: $120<Z_{tot}<180$ and $0.8<p_{\|,tot}/p_{proj}<1.1$.
 The number of events fullfiling these conditions is equal $5.9 * 10^{6}$.

\begin{figure}
\begin{center}
\includegraphics[width=0.95\linewidth]{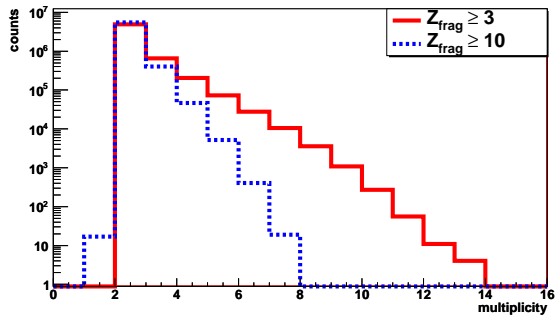}
\caption{
(color online) Multiplicity distributions of fragments with $Z_{frag}\geq3$ (red histogram) and $Z_{frag}\geq10$ (blue histogram), respectively.}
\label{fig07}
\end{center}
\end{figure}

For this class of well reconstructed events in the Au + Au reaction the multiplicity distribution of fragments with
charge $Z_{frag}\geq 3$ and $Z_{frag}\geq 10$ are
presented in Fig.~\ref{fig07}. One can notice here that the number of events with five or more fragments corresponding to above  charge thresholds
is equal about 116~000 and 6~000,
respectively.

\section{Data analysis}

Based on results of Ref. \cite{Skwira08}, \cite{Wilczynski10} and \cite{Wilczynski10a}, in analysis of ternary
and quaternary events one expects the observation of binary deep-inelastic collisions
followed by breakup of one or both primary reaction products.
Results of analysis of this particular class of reactions  can be found in \cite{Cap14}.

For the class of events with five fragments
one can consider at least two mechanisms responsible for the presence of the fifth heavy fragment:
(i) creation of the fragment in the interaction region (intermediate velocity source) for peripheral collisions
or (ii) the multifragmentation of the composite nuclear system formed in central collisions.

In order to investigate the reaction scenario responsible for events with five and more fragmets we have compared experimental
 data with ETNA and QMD model predicions. The ETNA model can simulate the decay of nuclear system assuming compact
 and noncompact freeze out configurations \cite{Sochocka09}.
In this model three freeze out configurations are considered: (i) ball geometry
 with volume 3 and 8 times greater than normal nuclear volume
$V_{0}$ (fragments uniformly distributed inside the sphere); (ii) fragments distributed on the surface
 of the sphere mentioned above (bubble configuration); (iii) fragments distributed on the ring
 with diameter 12 fm and 15 fm (toroidal configuration). In this model we consider  events corresponding to central
 collisions only (0-3 fm impact parameter range).

In order to simulate the contribution
 from noncentral collisions the QMD model \cite{Aichelin} calculations were performed in the full impact parameter range 0 - 12 fm.
 In our analysis the QMD code developed by Lukasik et al. \cite{QMD} was used. This code takes into account:
(i) protons and neutrons (in standard QMD each nucleon has an effective Z/A
charge); (ii) momentum dependent Pauli potential (Skyrme + Coulomb + Symmetry + Surface + Pauli) is used instead of
the Yukawa potential; (iii) initial nuclei in their
real ground states with minimum energy (thanks to the Pauli potential) are prepared,(iv) strict angular momentum conservation in collisions is applied;
(v) to simulate the Pauli blocking: in Lukasik code the overlap of 6-dimensional
Gaussians is used, unlike in standard QMD where overlap of appropriate spheres in configuration and momentum space is included.

\subsection{The shape sensitive observables}
In our analysis several observables sensitive to the freeze-out break-up configuration are investigated. As a first, we consider the shape
of events in the momentum space \cite{cugnon}. The diagonalization of the momentum tensor gives three
eigenvalues $\lambda_{i}$ and three eigenvectors $\overrightarrow{e_{i}}$.
 The sphericity and coplanarity variables are defined as:

 \begin{equation}
 s=1.5(1-\lambda_{1}),
 c=\frac{\sqrt{3}}{2}(\lambda_{2}-\lambda_{3}),
 \end{equation}

 where $\lambda_{1}> \lambda_{2}>\lambda_{3}$ are normalized to their sum.

\begin{figure}
\begin{center}
\includegraphics[width=\columnwidth]{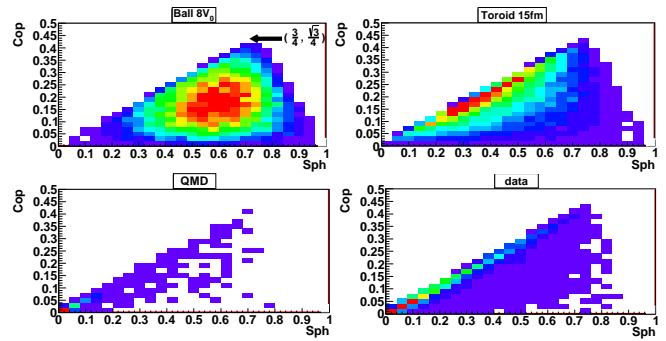}
\caption{(color online) The coplanarity vs sphericity distributions for Ball $8V_{0}$, Toroid 15 fm, QMD and experimental data.}
\label{fig08}
\end{center}
\end{figure}

In the coplanarity vs sphericity plane all events are located inside a triangle
defined by points (0,0), $(\frac{3}{4},\frac{\sqrt{3}}{4})$, and (1,0). In Fig.~\ref{fig08} the (s,c) distribution for experimental
 data is compared to the ETNA model predictions for Ball $8V_{0}$, Toroid 15 fm freezeout decay configurations and with QMD predictions.
In the case of ball geometry the maximum of the corresponding distribution is located in the centre
of the triangle. For toroidal configuration the distribution is located closer to
the line (0,0), $(\frac{3}{4},\frac{\sqrt{3}}{4})$. One can see that the experimental distribution looks very similar
to QMD distribution which is dominated by noncentral collisions contribution.

\begin{figure}
\begin{center}
\includegraphics[width=0.96\columnwidth]{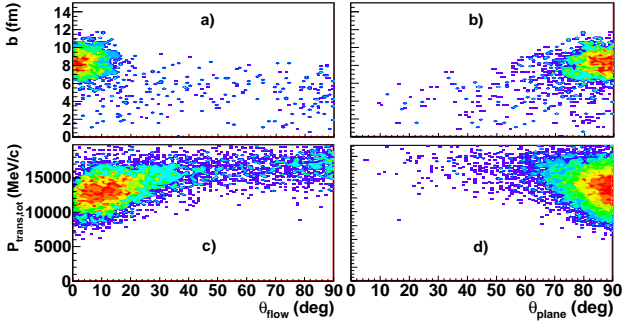}
\end{center}
\caption{(color online) The impact parameter vs $\theta_{flow}$ dependence for QMD model predictions (panel a)
and total transeverse momentum, $p_{trans}$, vs $\theta_{flow}$ dependence for experimental data (panel c).
The impact parameter vs $\theta_{plane}$ dependence for QMD model predictions (panel b)
and total transeverse momentum, $p_{trans}$, vs $\theta_{plane}$ dependence for experimental data (panel d).}

\label{fig09} 
\end{figure}

In order to reduce noncentral contribution we have investigated for the QMD model predictions
 the dependence between flow angle, $\theta_{flow}$, and impact parameter (see Fig.~\ref{fig09} (panel a), where
$\theta_{flow}$ is the angle between beam axis and the eigenvector $\overrightarrow{e_{1}}$
 for the largest eigenvalue $\lambda_{1}$.
 One can see on this plot that most noncentral events are located at small $\theta_{flow}$ angles.
The similar dependence is observed for experimental data between $\theta_{flow}$ and total transverse
 momentum, $p_{trans}$, used as impact parameter estimator (see Fig.~\ref{fig09} (panel c).
 We decide to reduce contribution of noncentral events both for experimental data and model predictions by using
 the condition $\theta_{flow} > 20^{o}$.

\begin{figure}
\begin{center}
\centerline{%
\includegraphics[width=1.\columnwidth]{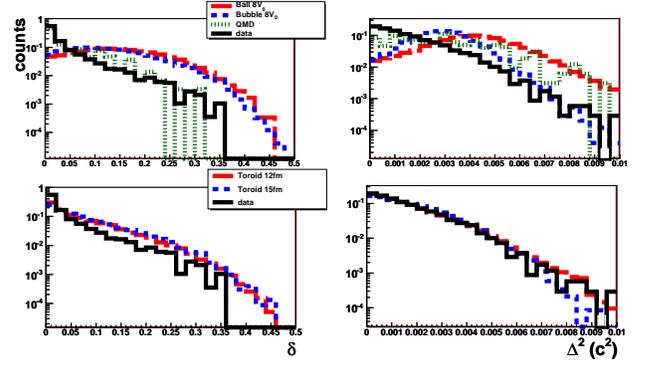}}
\caption{(color online) In the upper left panel the $\delta$ distributions are presented for experimental data, Ball 8$V_{0}$, Bubble 8$V_{0}$
freeze-out geometries and QMD predictions. In the bottom left panel the experimental distribution is
compared with predictions for Toroid 12 fm and Toroid 15 fm configurations.
In the right panels the $\Delta^{2}$ distributions for experimental data and model predictions are shown.
All the distributions presented here are constructed using the condition $Z_{frag}\geq 10$ and $\theta_{flow}>20^{o}$.}
\label{fig10}
\end{center}
\end{figure}

The $\delta$, and $\Delta^{2}$ observables as most sensitive to the shape of freeze out configurations were selected \cite{Sochocka09}.
The $\delta$ variable is related to sphericity and coplanarity variables .
The $\delta$ variable measures the distance between a given point of the (s,c) distribution
 and the line defined by points (0,0), $(\frac{3}{4},\frac{\sqrt{3}}{4})$.
In the Fig.~\ref{fig10} (left panels) the $\delta$ distributions are presented for experimental data, ETNA model predictions for
considered freeze-out geometries and QMD predictions. One can see here that the $\delta$ distribution for experimental data is similar
 to that corresponding QMD predictions. The biggest difference can be observed with the distribution for Ball $8V_{0}$ configuration.

The $\Delta^{2}$ variable used in our analysis gives a measure of the event flatness in the velocity space.
For each event we are establishing the
plane in the velocity space. The parameters of this plane are selected in the way that the sum of squares
of distances between the plane and the endpoints ($v_{x,i}, v_{y,i}, v_{z,i}$) of velocity vectors reach the minimum value.
 This last quantity is called the $\Delta^{2}$ parameter and is defined as:

\begin{equation}
\Delta^{2}=min[\sum_{i=1}^{N_{fr}}(d_{i}^{2}(A,B,C,D)],
\end{equation}

where:

\begin{equation}
d_{i}=\frac{|A\cdot v_{x,i}+B\cdot v_{y,i}+C\cdot v_{z,i}+D|}{\sqrt{A^{2}+B^{2}+C^{2}}},
\end{equation}

and parameters A, B, C, and D are the plane parameters. The plane parameters and
the velocities of fragments are in the velocity of light units.

The $\Delta^{2}$ distributions are shown in Fig.~\ref{fig10} (right panels) for data and model predictions.
One can see here that for $\Delta^{2}$ variable the biggest difference between experimental distribution and model predictions
is observed for the Ball 8$V_{0}$, and Bubble 8$V_{0}$ configurations. In contrast to that, the experimental data  seem to be more
consistent with the simulations assuming toroidal freeze-out configurations.

In relation with $\Delta^{2}$ parameter one can define an angle, $\theta_{plane}$, between the beam direction and vector normal
to the plane defined by parameters A, B, C, and D. For events corresponding to noncentral collisions, where most of reaction products
are located in the reaction plane, $\theta_{plane}$ should be close to $90^{o}$. This behavior is illustrated in Fig.~\ref{fig09} (panel b)
for QMD model predictions, where most of noncentral events are located in the reaction plane.
The similar dependence is observed for experimental data between $\theta_{flow}$ and total transverse
momentum, $p_{trans}$, used as impact parameter estimator (see panel d).

\begin{figure}
\begin{center}
\includegraphics[width=\linewidth]{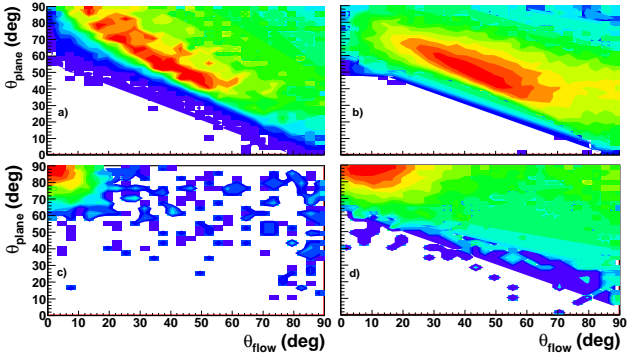}
\caption{
(color online) The dependence between $\theta_{plane}$ and $\theta_{flow}$
 for Ball $8V_{0}$ (panel a), Toroid 15 fm (panel b), QMD (panel c) and experimental data (panel d).}
\label{fig11}
\end{center}
\end{figure}

The dependence between $\theta_{plane}$ and $\theta_{flow}$
 for Ball $8V_{0}$, Toroid 15 fm, QMD and experimental data is presented in Fig. ~\ref{fig11}.
One observe here that for experimental data most of events is located in the region selected by conditions $\theta_{flow}<20^{o}$ and $\theta_{plane}>75^{o}$.
The same behavior is observed in the case of QMD calculations. These observations indicate that such events correspond to noncentral collisions.
For the Ball $8V_{0}$ configuration one observes the correlation between $\theta_{flow}$ and $\theta_{plane}$ angles.
For toroidal configuration the correlation between these angles is even stronger. Most of these events is located in the region
defined by conditions $\theta_{flow}>20^{o}$ and $\theta_{plane}<75^{o}$.

Following the method proposed in Ref. \cite{Sochocka09} we select events corresponding to a toroidal shape by the set of conditions:

 \begin{equation}
 \Delta^{2} <0.001~c^{2}~and~\delta <0.05 .
 \end{equation}

 As an efficiency measure of the above conditions we take ratio of number of events fulfilling
the selection conditions to the number of events with five and more heavy fragments.
Hereafter, this ratio is called the efficiency factor (EF).

\begin{figure}
\begin{center}
\includegraphics[width=0.9\linewidth]{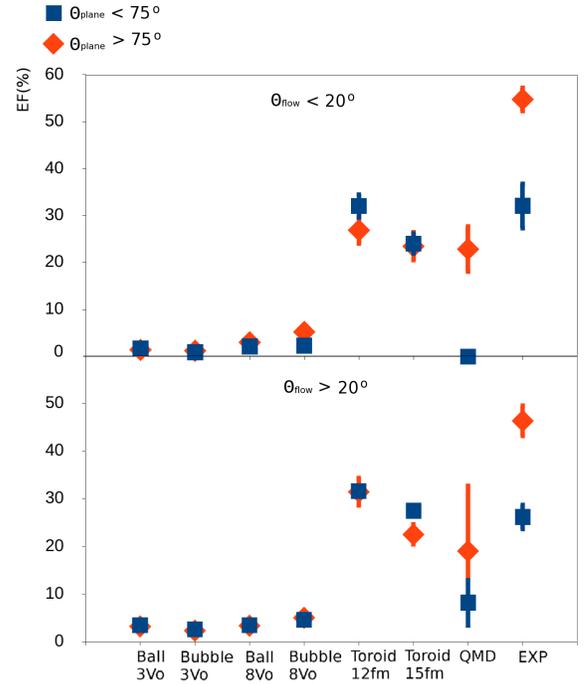}
\caption{
(color online) The EF values for different windows of $\theta_{plane}$ and $\theta_{flow}$. The presented results were sorted using the condition $Z_{frag}\geq 10$.
}
\label{fig12}
\end{center}
\end{figure}

The results of this procedure are presented in the Fig. ~\ref{fig12}
for different regions of $\theta_{flow}$ and $\theta_{plane}$ angles.
As one can see the EF is very low for spherical freeze-out configurations with respect to the corresponding values for
toroidal configurations.

For QMD calculations the value of the efficiency factor is strongly dependent on the $\theta_{plane}$ range.
The condition $\theta_{plane} < 75^{o}$ reduces the number of flat noncentral events
mostly located in the reaction plane.
For events selected additionally by the condition $\theta_{flow} < 20^{o}$ the EF drops to zero.

 For experimental data the value of the efficiency factor is about 50\% for events located in the reaction plane ($\theta_{plane} > 75^{o}$) and
is reduced by factor of 2 for events perpendicular to the reaction plane. These values are
weakly dependent on the $\theta_{flow}$ angle range.

One observes that the values of the EF for experimental data are much larger than the correspondig predictions for QMD model. The biggest
difference is observed for events located outside the reaction plane ($\theta_{plane} < 75^{o}$) at small $\theta_{flow}$ angles.

In order to investigate a possible formation of toroidal configurations in our analysis
we selected the region where according to ETNA predictions the toroidal configuration is most pronounced in the $\theta_{flow}$ and $\theta_{plane}$
plane ($\theta_{plane} < 75^{o}$ and $\theta_{flow} > 20^{o}$).
In Table ~\ref{table_1} the efficiency factor values are given for experimental data and model predictions.
 The efficiency factor values are shown for four threshold values of the fragment charge.

\begin{table}[hbt]
 \begin{center}
 \vspace{0.5cm}
 \begin{tabular}{|c|c|c|c|c|}
 \multicolumn {3}{c}{Efficiency factor (\%)} \\
 \hline
 Configuration & $Z_{frag}\geq 3$ & $Z_{frag}\geq 10$ & $Z_{frag}\geq 15$& $Z_{frag}\geq 20$\\
 \hline
 Ball $3V_{0}$ & 3.3 $\pm$ 0.2 & 3.5 $\pm$ 0.2 & 3.5 $\pm$ 0.2 & 3.5 $\pm$ 0.2 \\
 \hline
 Bubble $3V_{0}$ & 2.4 $\pm$ 0.2 & 2.6 $\pm$ 0.2 & 2.7 $\pm$ 0.2 & 2.7 $\pm$ 0.2\\
 \hline
 Ball $8V_{0}$ & 3.2 $\pm$ 0.2 & 3.5 $\pm$ 0.2 & 3.5 $\pm$ 0.2 & 3.5 $\pm$ 0.2\\
 \hline
 Bubble $8V_{0}$ & 3.9 $\pm$ 0.2 & 4.6 $\pm$ 0.2 & 4.7 $\pm$ 0.2 & 4.7 $\pm$ 0.2\\
 \hline
 Toroid 12 fm & 29.7 $\pm$ 0.6 & 31.6 $\pm$ 0.6 & 31.8 $\pm$ 0.6 & 31.9 $\pm$ 0.6\\
 \hline
 Toroid 15 fm & 25.2 $\pm$ 0.5 & 27.5 $\pm$ 0.5 & 27.7 $\pm$ 0.5 & 27.8 $\pm$ 0.5 \\
 \hline
 QMD & 13.7 $\pm$ 3.4 & 8.2 $\pm$ 4.7 & 6.3 $\pm$ 5.5 & N/A\\
 \hline
 data & 27.1 $\pm$ 0.7 & 26.2 $\pm$ 2.5 & 26.2 $\pm$ 4.8 & 21.1 $\pm$ 8.0\\
 \hline
 \end{tabular}
 \vspace {0.1cm}
 \caption{ \footnotesize
The efficiency factor at incident energy 23 AMeV for four threshold values of the fragment charge
for events selected by conditions $\theta_{flow}>20^{o}$ and $\theta_{plane}<75^{o}$.
}
\label{table_1}
\end{center}
\end{table}

 From Table ~\ref{table_1} we notice that the EF values for experimental data are very close to the model predictions for toroidal configurations.
 This observation may be one of arguments in favor of the formation of toroidal/flat freeze-out configuration
 created in the Au + Au collisions at 23 AMeV.

\subsection{Other observables}

In order to get additional evidence to support the hypothesis that toroidal objects are created the behaviour of other observables was investigated.
We consider here for each event separately:  (i) standard deviation of fragment mass ($\sigma_{A}$),
(ii) relative velocities of fragments pairs ($v_{ij}$), (iii) mean velocities of fragments as a function of their mass.

First we construct these observables
for events selected by conditions $\theta_{flow}>20^{o}$ and $\theta_{plane}<75^{o}$, where observation of toroidal freeze-out configurations is expected.
The distributions of these observables are generated for flat events selected by condition (7) (thick green histograms) and
non-flat events (thin red histograms) selected by condition:

\begin{equation}
\Delta^{2} >0.001~c^{2}~and~\delta >0.05.
\end{equation}

\begin{figure}
\begin{center}
\includegraphics[width=0.95\linewidth]{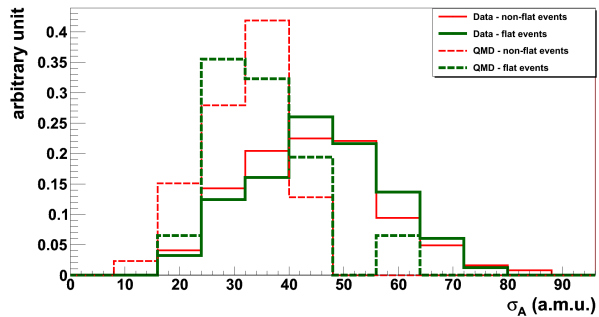}
\caption{(color online) The distributions of standard deviation of the fragment mass for non-flat events (red lines) and flat events (green lines)
for experimental data (solid lines) and QMD model predictions (dashed lines). All the distributions presented here are constructed using the condition $Z_{frag}\geq 10$.}
\label{fig13}
\end{center}
\end{figure}

Comparison of the $\sigma_{A}$ distributions (Fig.~\ref{fig13}) for flat and non-flat events indicates
that in the case of flat events this distribution is slightly shifted to larger values.
This observation is in contrast with the expectation that for the flat events the
enhanced similarity in the size of fragments should be visible. The corresponding distributions for QMD calculations are similar (dashed lines).
Their centroids are shifted to smaller values with respect to experimental data.

\begin{figure}
\begin{center}
\includegraphics[width=0.95\linewidth]{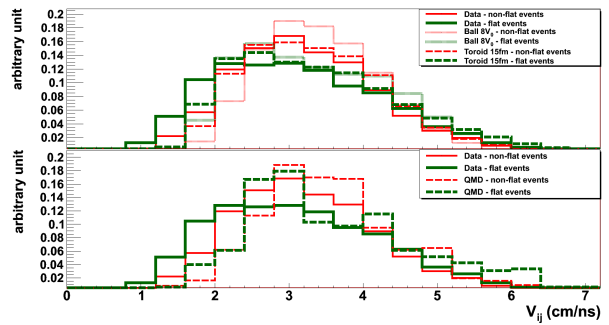}
\caption{(color online) The distribution of relative velocities $v_{ij}$ of fragments pairs for non-flat events (red lines) and flat events (green lines)
for experimental data (solid lines) compared with: Ball 8$V_{0}$ and Toroid 15 fm (upper panel) and QMD model predictions (bottom panel).
All the distributions presented here are constructed using the condition $Z_{frag}\geq 10$.}
\label{fig14}
\end{center}
\end{figure}

In Fig.~\ref{fig14} one observes that the distribution of relative velocities for flat events is shifted to smaller
velocities in respect to non-flat events.
The corresponding distributions for Toroid 15 fm and Ball 8$V_{0}$ ETNA model predictions show a similar dependence.
This observation may indicate that the behaviour of these $v_{ij}$ distributions is insensitive to the shape
of the freeze-out configuration.

\begin{figure}
\begin{center}
\includegraphics[width=0.95\linewidth]{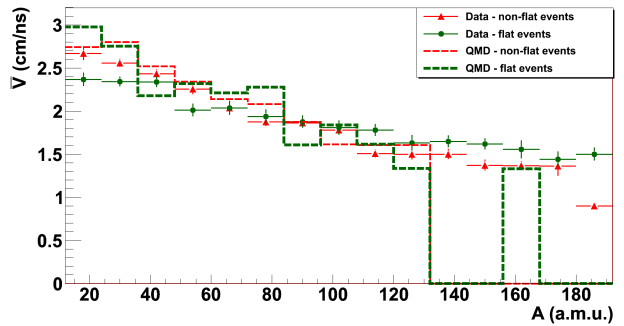}
\caption{(color online) The distributions of mean velocities of fragments as a function of their mass for non-flat events (red lines) and flat events (green lines)
for experimental data (points with error bars) and QMD model predictions (dashed lines). All the distributions presented here are constructed using the condition $Z_{frag}\geq 10$}
\label{fig15}
\end{center}
\end{figure}

In Fig.~\ref{fig15} the distributions of mean velocities of fragments
as a function of their mass for a flat and non-flat events are presented. On can observe that for flat events velocities
of fragments decrease weaker with mass comparing to the same dependence for non-flat events.
Comparison with same dependences presented for Pb + Ag and Pb + Au systems at 29 AMeV \cite{Jouault}
indicates that toroidal configurations may by created for some subclass of flat events.

\begin{minipage}[t]{0.45\textwidth}
    \begin{center}
        \begin{adjustbox}{center, width=0.8\columnwidth-10pt}

			\begin{tabular}{|c|c|c|c|c|c|}
			\hline
			Observable & threshold & $\theta_{flow}>20^{\circ}$ & $\theta_{flow}>20^{\circ}$ & $\theta_{flow}<20^{\circ}$ & $\theta_{flow}<20^{\circ}$ \\ 
			& & $\theta_{plane}<75^{\circ}$ & $\theta_{plane}>75^{\circ}$ & $\theta_{plane}<75^{\circ}$ & $\theta_{plane}>75^{\circ}$ \\
			\hline \hline
			\multirow{5}{*}{ $\sigma_{A} (a.m.u.)$}
			& $Z_{frag}\geq 3$ & 72.09 $\pm 0.47 $& 71.09 $\pm 0.35$ & 76.43$\pm 0.52$ & 73.33$\pm 0.13$ \\
			& $Z_{frag}\geq 10$ & 47.01 $\pm 1.88 $& 47.15 $\pm 1.33$ & 46.41 $\pm 2.68$ & 45.06 $\pm 0.58$ \\
			& $Z_{frag}\geq 15$ & 38.31 $\pm 2.98 $ & 38.53 $\pm 1.24$ & 35.24 $\pm 5.11$ & 35.58 $\pm 0.98$ \\
			& $Z_{frag}\geq 20$ & 31.01$\pm 6.68 $ & 31.10 $\pm 2.60$ & 25.15 $\pm 5.15$ & 27.17 $\pm 1.59$ \\
			& $Z_{frag}\geq 25$ & 17.51$\pm 5.07 $ & 18.95 $\pm 5.86$ & 20.94 $\pm 4.82$ & 18.50 $\pm 2.23$ \\ 
			\hline
			\multirow{5}{*}{$v_{ij} (cm/ns)$}
			& $Z_{frag}\geq 3$ & 3.01 $\pm 0.01 $ & 3.17 $\pm 0.01$ & 3.27 $\pm 0.02$ & 3.36 $\pm 0.01$\\
			& $Z_{frag}\geq 10$ & 3.13 $\pm 0.05 $ & 3.30$\pm 0.03$ & 3.30 $\pm 0.08$ & 3.51 $\pm 0.02$\\
			& $Z_{frag}\geq 15$ & 3.16 $\pm 0.08 $ & 3.27 $\pm 0.05$ & 3.27 $\pm 0.15$ & 3.49 $\pm 0.04$\\
			& $Z_{frag}\geq 20$ & 3.14 $\pm 0.24 $ & 3.25$\pm 0.11$ & 3.24 $\pm 0.52$ & 3.50 $\pm 0.06$\\
			& $Z_{frag}\geq 25$ & 2.98 $\pm 0.31 $ & 3.28 $\pm 0.33$ & 3.26 $\pm 0.81$ & 3.46 $\pm 0.13$\\
			\hline
			\end{tabular}
        \end{adjustbox}
        \captionof{table}{\footnotesize The mean values of mass standard deviation of the fragments, and of relative velocities $v_{ij}$ of fragments pairs
		for flat events located in different windows of $\theta_{flow}$ and $\theta_{plane}$ angles.}
		\label{table_2}
    \end{center}
\end{minipage}

\begin{figure}
\begin{center}
\includegraphics[width=0.95\linewidth]{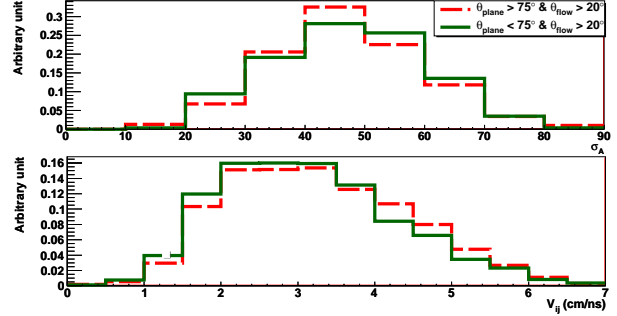}
\caption{(color online) The distributions of standard deviation of the fragment mass (upper panel), and the distributions of relative velocities $v_{ij}$ of fragments
pairs (bottom panel) for flat events.
The red dashed lines corresponds to events located  inside the reaction plane and the green solid lines correspond
to events located outside the reaction plane. All the distributions presented here are constructed using the condition $Z_{frag}\geq 10$}
\label{fig16}
\end{center}
\end{figure}

Properties of flat events in the region where observation of toroidal freeze-out configurations is expected ($\theta_{flow}>20^{o}$ and $\theta_{plane}<75^{o}$)
can be also compared with properties of flat events corresponding to other regions of $\theta_{flow}$ and $\theta_{plane}$ angles. Here the considered regions are the same
as presented in Fig. ~\ref{fig12}.
The distributions for $\sigma_{A}$ of fragments, and $v_{ij}$ of fragments pairs are presented in Fig. ~\ref{fig16} using the condition $Z_{frag}\geq 10$.
The mean values
of these distributions are listed in Table ~\ref{table_2}.
We can notice here that the corresponding mean values of the  distribution of $\sigma_{A}$ are similar for all $\theta_{flow}$ and
$\theta_{plane}$  windows for a given threshold value of the fragment charge $Z_{frag}$.
Such observation shows us that information carried by $\sigma_{A}$ can  not be used as an indication of
toroidal objects formation. For $v_{ij}$ distributions one observes that the mean values for class of events located outside the reaction plane
are smaller in comparison to the case of events located in the reaction plane. The smallest mean values are seen
for the region where observation of toroidal freeze-out configurations are expected.
 This observation may be used as an indication that for events located outside the reaction plane freeze-out configuration is more extended in comparison with
that for events located inside reaction plane.

Results obtained for the considered observables suggest that the formation of toroidal configurations can be related to a fraction
of flat events tilted with respect to the reaction plane ($\theta_{plane} < 75^{o}$).
The probability for these events is much greater than the prediction of the QMD model. The nature of these events should be investigated.

Assuming that the total number of detected events corresponds to 80\% of total reaction cross section, the cross section related to creation of
flat tilted events located in the region where observation of toroidal freeze-out configurations is expected can be estimated to be equal $17 \mu b$.

\section{Summary}
We presented an analysis of events produced in Au + Au collisions at
23 AMeV. Basic information about data calibration procedure were summarized.
The bulk properties of the experimental data were shown. The experimental data were compared with ETNA
and QMD model predictions. Proximity of efficiency factor values for experimental data and toroidal freeze-out configurations  may be used
 as an indication of the formation of an exotic freeze-out configuration.
The juxtaposition of the standard deviation of fragment
mass values for events located outside and inside the reaction plane are not suggestive of a toroidal freeze-out configuration.
Comparison of distributions of relative velocities   for event with different orientation in respect to reaction plane
gives evidence that the freeze-out configuration is more extended for events located outside reaction plane.
The behavior of mean velocities of fragments as a function of their mass for flat and non-flat events gives an indication that toroidal configuration
may be created for some subclass of flat events.

The probability of apperence of these flat events
is much greater than the prediction of the QMD model. The nature  of flat events tilted with respect to the reaction plane should be investigated.

\section{Acknowledgments}

Special thanks are due the INFN-LNS operating crew for providind an excellent beam.
This work has been partly supported by the National Science Centre of Poland (grant N N202 180638, 2013/09/N/ST2/04383)).

\end{document}